\documentclass{emulateapj}
\slugcomment{{\sc Accepted to ApJ:} 2011, December 17} 
\usepackage{color}
\shorttitle{}
\shortauthors{Sharon et al.}

\newcommand{\magA}{$25.1 ^{+3.2}_{-2.5}$}
\newcommand{\magtotal}{$28.4 ^{+3.4}_{-2.7}$}
\newcommand{\magslit}{$42.2 \pm 5.5$}
\newcommand{\magAa}{$10.4 ^{+1.1}_{-0.8}$}
\newcommand{\magAb}{$20.6 ^{+2.6}_{-2.2}$}
\newcommand{\magAc}{$9.7 ^{+1.1}_{-0.9}$}
\newcommand{\magC}{$3.0 ^{+0.2}_{-0.1}$}
\newcommand{\magfifth}{$0.38 ^{+0.06}_{-0.12}$}
\newcommand{\SFRinslit}{$ 29 \pm   8$}

\newcommand{\Px}{$-2.23 ^{+0.14}_{-0.20}$}
\newcommand{\Py}{$-2.03 ^{+0.12}_{-0.13}$}
\newcommand{\Pe}{$0.40 ^{+0.03}_{-0.02}$}
\newcommand{\Ptheta}{$9.1 ^{+0.5}_{-0.4}$}
\newcommand{\Prc}{$ 55 ^{+  5}_{-  4}$}
\newcommand{\Psigma}{$1139 ^{+ 17}_{- 14}$}
\newcommand{\Psigmab}{$258 ^{+ 38}_{- 29}$}
\newcommand{\zSseven}{$2.29 ^{+0.08}_{-0.05}$}
\newcommand{\Pxb}{$0.03$ [$-0.05$ $0.05$]}
\newcommand{\Pyb}{$-0.02$ [$-0.05$ $0.05$]}
\newcommand{\Peb}{$0.19$ [$0.00$ $0.30$]}
\newcommand{\Pthetab}{$153$ [$  0$ $180$]}
\newcommand{\Prcb}{$0.5$ [$0.0$ $2.0$]}
\newcommand{\Psigmac}{$ 85$ [$ 10$ $100$]}
\newcommand{\Prcutc}{$  6$ [$  1$ $ 25$]}
\newcommand{\Psigmad}{$ 86$ [$ 10$ $100$]}
\newcommand{\Prcutd}{$  3$ [$  1$ $ 25$]}

\newcommand{\arcname}{RCSGA 032727-132609}

\begin{document}

\title{Source Plane Reconstruction of The Bright Lensed Galaxy RCSGA 032727-132609\altaffilmark{*}}
\altaffiltext{*}{Based on observations made with the NASA/ESA 
  {\it Hubble Space Telescope}, obtained at the Space Telescope Science
  Institute, which is operated by the Association of Universities for
  Research in Astronomy, Inc., under NASA contract NAS 5-26555. These
  observations are associated with program GO-12267}

\author{Keren Sharon\altaffilmark{1}, 
{Michael D. Gladders\altaffilmark{1,2}}, 
{Jane R. Rigby\altaffilmark{3}}, 
{Eva Wuyts\altaffilmark{1,2,4}},
{Benjamin P. Koester\altaffilmark{2}}, 
{Matthew B. Bayliss\altaffilmark{1,2}},
{L. Felipe Barrientos\altaffilmark{5}}
}

\altaffiltext{1}{Kavli Institute for Cosmological Physics, University of Chicago, 5640 South Ellis Avenue, Chicago, IL 60637, USA.}
\altaffiltext{2}{Department of Astronomy and Astrophysics, University of Chicago, 5640 South Ellis Avenue, Chicago, IL 60637, USA}
\altaffiltext{3}{Observational Cosmology Lab, NASA Goddard Space Flight Center, Greenbelt MD 20771}
\altaffiltext{4}{Observatories of the Carnegie Institution of Washington, Pasadena, CA 91101, USA}
\altaffiltext{5}{Departamento de Astronomía y Astrofísica, Pontificia Universidad Católica de Chile, Avda. Vicuña Mackenna 4860, Casilla 306, Santiago 22, Chile}

\email{kerens@kicp.uchicago.edu}

\begin{abstract}
We present new {\it HST}/WFC3 imaging data of \arcname, a bright lensed galaxy at $z$=1.7 that is magnified and stretched by the lensing cluster RCS2 032727-132623. Using this new high-resolution imaging, we modify our previous lens model (which was based on ground-based data) to fully understand the lensing geometry, and use it to reconstruct the lensed galaxy in the source plane. This giant arc represents a unique opportunity to peer into 100-pc scale structures in a high-redshift galaxy. {This new source reconstruction will be crucial for a future analysis of the spatially-resolved rest-UV and rest-optical spectra of the brightest parts of the arc.}
\end{abstract}


\section{Introduction}
The study of galaxy formation and evolution relies on imaging and spectroscopy of high-redshift galaxies. The relevant data are typically limited by the surface brightness of these galaxies, and their spatial size only allows a measurement of the galaxy as a whole, limiting our ability to understand where in the galaxy processes like star formation occur.
To reach sufficient levels of signal-to-noise, studies often stack spectra of many galaxies, thus averaging over a population (e.g., Shapley 2003, Reddy et al. 2010).
The only means to study individual high-$z$ galaxies in detail, at least until the era of JWST and 30m-class telescopes, is to analyze lensed galaxies whose brightnesses are magnified, and spatial sizes are stretched by the gravitational potential of the lens, making them suitable for spatially-resolved spectroscopy. Such a rare case of a lensed galaxy was recently discovered in the second Red Sequence Cluster Survey (RCS2; Gilbank et al. 2011) Giant Arc Survey (Bayliss et al., in prep.), as an extremely bright, highly-stretched lensed galaxy at $z=1.7$, lensed by the cluster RCS2 032727-132623. The giant arc, labeled RCSGA 032727-132609, is $\sim 38\arcsec$ long, and has a moderately-magnified counter image. 
 In Wuyts et al. (2010) we report the discovery of RCSGA 032727-132609, spectral confirmation, the measurement and modeling of its spectral energy distribution from ground-based data, and a preliminary strong-lensing mass model of the lensing cluster. In Rigby et al. (2011; hereafter R11), we analyze a Keck/NIRSPEC spectrum of the brightest part of the arc to determine the physical conditions in this galaxy.
The interpretation of our findings in both papers relied on a preliminary lens model based on ground-based imaging, which did not allow unique identification of arc substructure at the level required for a robust lens model; this limited our ability to fully understand the lensed, highly distorted arc in terms of its source morphology. It also compromises the precision and accuracy with which we can determine the average magnification of the galaxy, a property crucial for calculating intrinsic properties such as the star formation rate.
In this paper, we use Hubble Space Telescope ({\it HST}) imaging with Wide Field Camera 3 (WFC3) to uniquely identify features in the lensed galaxy. With the dramatic improvement in spatial resolution (less then $0\farcs1$ FWHM in the UVIS channel) relative to the ground-based data ($0\farcs5$ seeing at best, see Wuyts et al. 2010), these data, along with newly-discovered additional multiply-lensed galaxies, are used to generate a robust lens model of the cluster. 
The observations and data reduction are described in \S~\ref{s.observations}. Lensing analysis and derived magnifications are presented in \S~\ref{s.lensing} and \S~\ref{s.magnification}; and in \S~\ref{s.spec}, the physical conditions of the source are revisited, based on the new lens model.
We assume a flat cosmology with $\Omega_{\Lambda} = 0.7$, $\Omega_{m} =0.3$, and $H_0 = 70$ km s$^{-1}$ Mpc$^{-1}$. Magnitudes are reported in the AB system.

\begin{figure*}
\epsscale{1.16}
\plotone{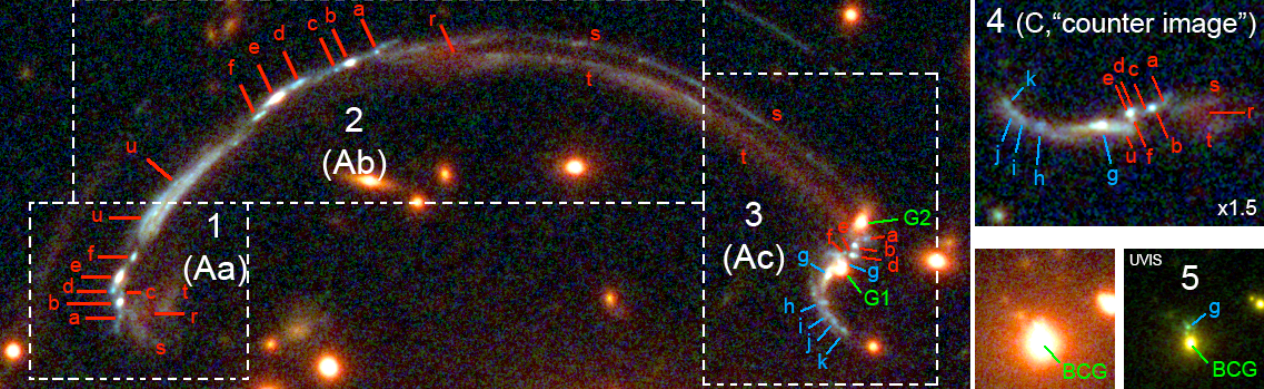}
\caption{Identification of substructure in the five images of the multiply-lensed source \arcname. The color rendition is composed of F160W, F125W, F098M (red); F814W, F606W (green); and F390W (blue), to highlight color gradients in the arc. The dashed lines approximately enclose the parts of the arc that compose each image, indicated by numbers. The labels used in Wuyts et al. (2010) are also indicated for clarity. We include another color rendition for image 5, composed of the UVIS bands only (F814W, F606W, F390W) and different scaling, to better separate image 5 from the BCG.
The emission knots in images 1-4 are labeled with letters a through k. The BCG is labeled in the renditions of image 5, as are the two cluster galaxies (G1, G2) that interrupt image 3. Notice that the a--f knots (red labels) appear in all of the images (although hard to identify in image 5), while the brightest knot of the source (labeled g) and the ``blue'' arm (h--k; blue labels) only appear in images 3, 4 and 5 (see text for an explanation). In image 3, the lensing perturbation by a cluster galaxy results in another instance of some parts of the source, likely the emission knot labeled g.
We note that knots c, d, and f in images 3 and 4 were not used as constraints in the lens modeling, since their positions are not as well-determined as in the other images. The very red extension at the end of the ``red'' arm (labeled r,s,t), and the highly-magnified region labeled u, were also not set as positional constraints in the model, but were used to evaluate models by demanding an agreement between model-predicted source morphologies derived from the different images.} \label{fig.1}
\end{figure*}

\section{Observations}\label{s.observations}
The field of RCS2 032727-132623 was imaged by {\it HST}/WFC3 on 2011, March 1 UT, during four orbits, as part of GO program 12267 (PI: Rigby). The data were acquired using six broad-band filters and one narrow-band filter: 
F390W (total exposure time 1401 s), F606W (1003 s), and F814W (2133 s) in the UVIS channel, and F098M (1212 s), F125W (862 s), F132N (2212 s), and F160W (862 s) in the IR channel. 
The IR medium/wide filters were chosen in order to measure the Balmer break, and to subtract the continuum light from the narrow-band IR filter, F132N, which samples the H$\beta$ line at 1.3153$\mu$m. 
The UVIS imaging samples the spatially-resolved UV spectral slope.

The imaging strategy consists of four sub-pixel dither positions in each filter to reconstruct the PSF, reject cosmic rays, and compensate for chip-gap. The IR-channel exposures were obtained in the SPARS50 read-out sequence mode. 
Individual frames were processed with the standard WFC3 calibration pipeline, and combined using the Multidrizzle routine
(Koekemoer et al. 2002) to remove cosmic-ray hits and dead or hot pixels, with a square kernel, \texttt{pixfrac$=$1.0} and \texttt{scale$=$INDEF}. 

Since the pixel scale and the point spread function (PSF) varies between filters, a set of PSF-matched images was created by convolving each image with a gaussian kernel to approximate the PSF of the F160W image, which has the broadest PSF. These images were used for measurements of magnitudes and colors. For the purpose of detection and identification of multiple images, the unconvolved, best-resolution images were used, and are shown here.

\begin{figure*}
\epsscale{1.16}
\plotone{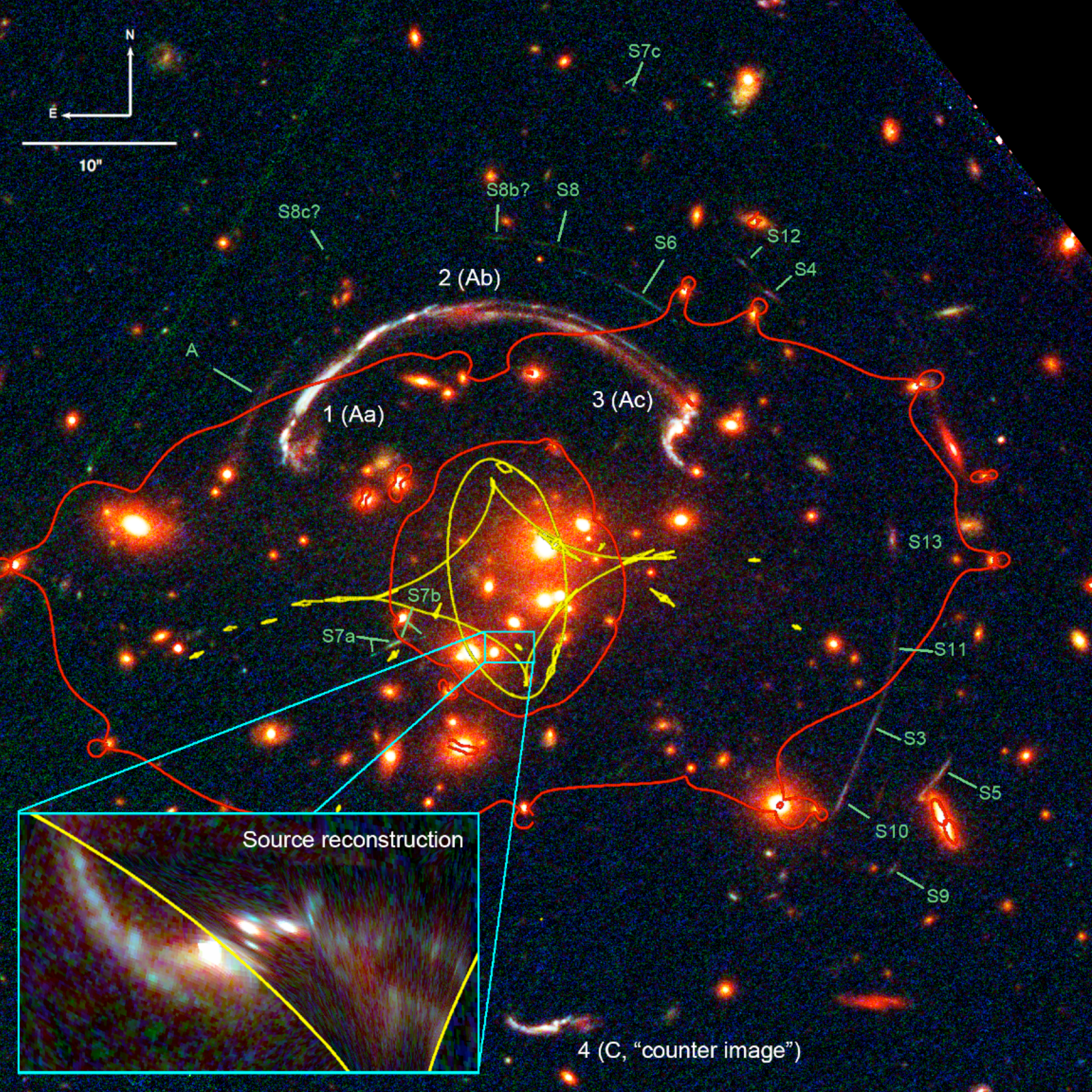}
\caption{{\it HST}/WFC3 image of the core of RCS2 032727-132623. The details of the color rendition are the same as in Figure~\ref{fig.1}. The critical curves of the best-fit model are overplotted in red, and the source-plane caustics are in yellow. The un-lensed size of the source and its location in the source plane are indicated by the cyan rectangle close to the center of the cluster. The inset shows a reconstruction of the source. The giant arc and additional arc candidates are labeled; we keep the same notation as in Wuyts et al. (2010).\label{fig.2}}
\end{figure*}

\section{Strong Lensing Mass Model}\label{s.lensing}
\subsection{Identification of Multiply-Imaged Galaxies}
\subsubsection{RCSGA 032727-132609}
The source galaxy of RCSGA 032727-132609 is lensed into five images, as labeled in Figure~\ref{fig.1}: three of them are merged, forming the giant arc; a fourth image appears as the ``counter image'' $31.5\arcsec$ south of the brightest cluster galaxy (BCG); a fifth, demagnified, image is predicted by the lens model, and a careful inspection of the {\it HST} imaging reveals its presence $0\farcs6$ north of the BCG, partially embedded in the BCG light. 
The ground-based data that were used in Wuyts et al. (2010) already indicate the existence of substructure in the source galaxy and subsequently its lensed images. However, since those data lack the needed resolution, we were unable to uniquely match features in the different instances of the lensed galaxy with each other. Moreover, substructure in the ``counter image'' is not resolved in those data. The new {\it HST} data, combined with an initial lens model, make the interpretation of the morphology unambiguous.  Figure \ref{fig.1} labels the different emission knots in each instance of the multiply-lensed galaxy. Notice that not all the parts of the source galaxy are represented in all of its lensed instances (see \S~\ref{s.source}).
The coordinates of each emission knot were used as constraints in computing the lens model.
\subsubsection{Other Background Sources}\label{s.arcs}
Apart from the extraordinary giant arc, several other background galaxies appear to be lensed into arcs, and are labeled in Figure~\ref{fig.2}. We use the lens model (\S~\ref{s.model}) to interpret the lensing nature of these secondary arcs, and estimate the redshifts of some of them, which are briefly discussed here. {The redshifts are estimated by using the model to predict the possible locations of the secondary arcs for different redshifts, and ruling out redshift ranges for which the predicted image configuration is not in line with the observed one. }
  Where appropriate, we use arc identifications from Wuyts et al. (2010) below and label them in Figure~\ref{fig.2}.

\noindent (A) A faint tangential arc appears $2\farcs5$ outwards of the East component of the giant arc. Our lensing interpretation suggests that it is a part of the same source galaxy of the giant arc (see \S~\ref{s.source}). {Since other interpretations are possible, it is not used as constraint in the modeling process.}

\noindent (S7) A radial arc, detected $11\farcs5$ ESE of the BCG (labeled S7a and S7b), with a counter image predicted by the model and detected $\sim30\arcsec$ North of the BCG (S7c), with a good morphological resemblance, correct parity, and colors indistinguishable from those of S7a and S7b. Since the identification of all the multiply-lensed images of S7 is robust, we include this arc as a constraint in our lens modeling below, setting its redshift as a free parameter, allowing for it to be solved in the modeling process. We note that the {lens-}model-predicted redshifts of other secondary arcs (below) somewhat depend on the best-fit redshift of S7, \zSseven.

\noindent (S6) A merging pair $17\farcs5$ NNE of the BCG, estimated to be at $z\sim 2.2$. {It is not used as a constraint in the model.}

\noindent (S8) A greenish arc $19\farcs7$ north of the BCG, estimated to be at $z\gtrsim3.5$. The redshift estimate is supported by the galaxy colors; in particular, the galaxy emission is only detectable in bands redder than F390W. { This source is also not currently included as a constraint.}
\begin{deluxetable*}{lccccccc} 
 \tablecolumns{8} 
\tablecaption{Best-fit lens model parameters  \label{t.parameters}} 
\tablehead{\colhead{Halo }   & 
            \colhead{RA}     & 
            \colhead{Dec}    & 
            \colhead{$e$}    & 
            \colhead{$\theta$}       & 
            \colhead{$r_{\rm core}$} &  
            \colhead{$r_{\rm cut}$}  &  
            \colhead{$\sigma_0$}\\ 
            \colhead{(PIEMD)}   & 
            \colhead{($\arcsec$)}     & 
            \colhead{($\arcsec$)}     & 
            \colhead{}    & 
            \colhead{(deg)}       & 
            \colhead{(kpc)} &  
            \colhead{(kpc)}  &  
            \colhead{(km s$^{-1}$)}  } 
\startdata 
Cluster  & \Px   & \Py   & \Pe   & \Ptheta    & \Prc   & [2000] & \Psigma  \\ 
BCG       & \Pxb & \Pyb & \Peb & \Pthetab  & \Prcb & [40]     & \Psigmab  \\ 
Gal~1    & [9.426] &[9.306] &[0.54] &[63.3] &[0.08] &\Prcutc & \Psigmac \\ 
Gal~2    & [8.684] &[7.587] &[0.22] &[163.4] &[0.004] &\Prcutd & \Psigmad \\ 
L* galaxy  & \nodata & \nodata & \nodata & \nodata &  [0.15]  &     30 [0 70]&  120 [50 150]  \\ 
\enddata 
 \tablecomments{All coordinates are measured in arcseconds relative to the center of the BCG, at [RA, Dec]=[51.863302, -13.439714]. The ellipticity is expressed as $e=(a^2-b^2)/(a^2+b^2)$. $\theta$ is measured north of West. Error bars correspond to 1 $\sigma$ confidence level as inferred from the MCMC optimization. Values in square brackets are for parameters that are not optimized. Errors in square brackets represent the lower and upper limits that were set as prior in the optimization process, for parameters that were not well-constrained by the data. 
The location and the ellipticity of the matter clumps associated with the  cluster galaxies were kept fixed according to their light distribution.}
\end{deluxetable*} 

\subsection{Lens Model}\label{s.model}
A lens model was computed using the publicly-available software \texttt{LENSTOOL} (Jullo et al. 2007), using Monte Carlo Markov Chain (MCMC) optimization.
The cluster is represented by a pseudo-isothermal ellipsoid mass distribution (PIEMD; Limousin et al. 2005)\footnote{This profile is formally the same as dual Pseudo Isothermal Elliptical Mass Distribution (dPIE, see El\'iasd\'ottir et al. 2007).}, parameterized by its position $x$, $y$; a fiducial velocity dispersion $\sigma$; a core radius $r_{core}$; a scale radius $r_s$; ellipticity $e=(a^2−-b^2)/(a^2+b^2)$, where $a$ and $b$ are the semi major and semi minor axes, respectively; and a position angle $\theta$. All the parameters of the mass distribution are allowed to vary within priors. We set a gaussian prior on $\sigma$ based on the observed velocity dispersion of the cluster (Wuyts et al. 2010). Solving the equations in  El\'iasd\'ottir et al. (2007; Appendix A) for typical cluster values of $r_{core}$ and $r_s$, and the radius in which the velocity dispersion was measured, we find that a small scaling of $\sim 1\%$ is needed in order to convert the observed velocity dispersion to the PIEMD $\sigma$ parameter. 
Cluster galaxies were selected by their $V-I$ color with respect to the cluster red sequence in a color-magnitude diagram. We impose a magnitude cutoff, and include only galaxies brighter than $24$ mag in the $i$-band within $54\arcsec$ from the cluster center. This conservative selection guarantees that we take into account any red-sequence galaxy whose lensing contribution is large enough to deflect the lensed images by an amount comparable to the measurement error (see e.g.,  Jullo et al. 2007, Limousin et al. 2010). Each galaxy is represented by a PIEMD, with parameters that follow the observed properties of the galaxies through scaling relations (see Limousin et al. 2007 for description of the scaling relations). The free parameters, representative of an $L^*$ elliptical galaxy, are $\sigma^*$, which is allowed to vary in the range [50, 150] km s$^{-1}$, and $r_{cut}^*$ which is forced to be smaller than 70 kpc, following Limousin et al. (2008). 

The BCG is represented by a PIEMD, with $x$, $y$, $e$, and position angle fixed at the observed values, and $r_{cut}$ fixed at 40 kpc. 
{Two cluster galaxies (labeled G1 and G2 in Figure~\ref{fig.1}) are superimposed on the western portion of the giant arc (see also Wuyts et al. 2010). 
We include these galaxy individually, with $x$, $y$, $e$, and position angle fixed at the observed values.}
Although perturbation from cluster galaxies is an essential component in the model, the proximity of the observed arcs to cluster galaxies is not small enough to constrain their model parameters, meaning that a large region in the galaxy parameter space is allowed by the constraints. We therefore report the best-fit value, and the range that was set as prior instead of a formal uncertainty. 

{The best-fit model is found in an iterative manner. We start with an optimization in the source plane, allowing all the parameters to vary as described above. The resulting best-fit model is used as an initial guess in the final modeling process, using image-plane optimization. Since the cluster-member parameters are not constrained by the observed data, we fix them at  $\sigma^*=120$  km s$^{-1}$, and $r_{cut}^*=30$ kpc. The best-fit model has an image-plane RMS of $0.18 \arcsec$.}
The model parameters, best-fit values, and their uncertainties or priors, are enumerated in Table~\ref{t.parameters}.

In addition to the emission knots in the giant arc and its counter images, we use as constraints the positions of the radial arc (S7a, S7b; see S\ref{s.arcs}) at an unknown redshift. An initial lens model predicted a counter image $\sim30\arcsec$ North of the BCG (S7c), which is detected in the data with excellent morphological resemblance to the model-predicted image. We thus include the three images of S7 (S7a and S7b, which form a merging-pair radial arc, and S7c) as additional constraints. We set a prior on the redshift of this source to be larger than 1.7, as its critical curve, which is tightly constrained by the two arcs that form the radial arc, indicates that it is at a higher redshift than the primary giant arc.  
The modeling process predicts a redshift of \zSseven~ for this source. 
By including the three images of S7 in the model we are able to significantly reduce the uncertainty in model parameters, and consequently in the magnifications, from $\sim50\%$ to $\sim 10\%$. 
{Detailed simulations, in which we model the cluster assuming that the redshift of S7 is known}, indicate that securing a spectroscopic redshift for S7 would further reduce the magnification uncertainty to the few-percent level. This agrees with the results of Jullo et al. (2007), who tested simulated lens potential parameter degeneracies, and with Miralda-Escude (1995), who showed that radial arcs and their counter images provide a stringent constraint on the profile shape as well as the enclosed mass. Thus, while S7 already improves the magnification uncertainties by a factor of five (relative to a model constrained only by the primary giant arc and the counter image), measuring a redshift for it is an obvious high priority, since it would further improve the model uncertainties to $<10\%$.

{In total, we have 44 constraints and 19 free parameters}, resulting in an over-constrained model. 
In the future, one might be able to include additional sub-structure in the cluster mass distribution, to make the ultimate lens model.  For now, we choose to not include substructure whose existence is not directly motivated by an observed counterpart. 

\begin{figure*}
\epsscale{1.16}
\plotone{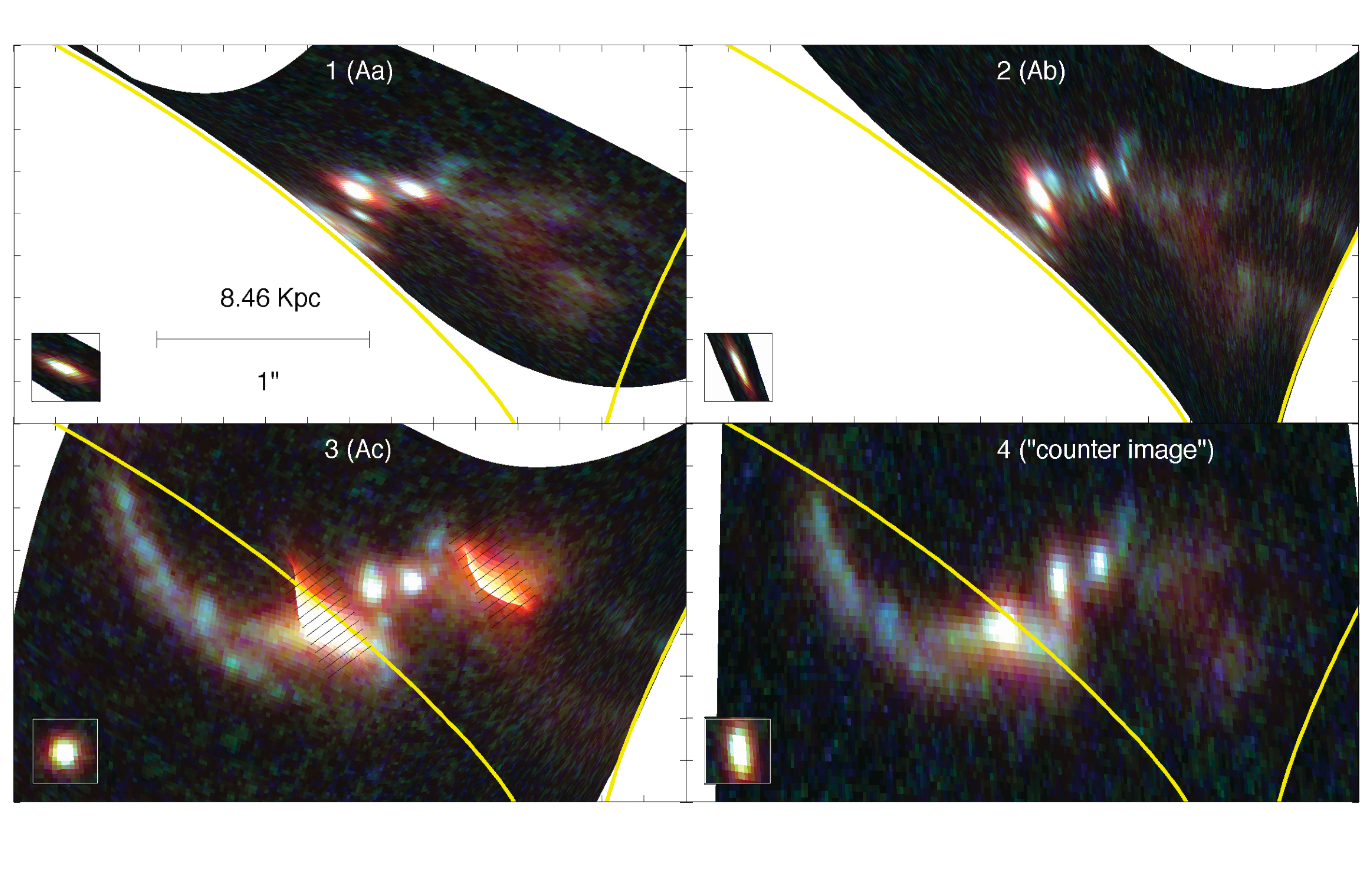}
\caption{Source reconstruction from each of the magnified images of \arcname. The source plane caustic is overplotted in yellow, and a scale of $1 \arcsec$, which corresponds to 8.46 Kpc at the source redshift, is indicated. The tick marks are 5 pixels apart with a pixel scale of 0.0396$\arcsec$ pixel$^{-1}$. The two cluster galaxies that are superimposed on image 3 (Ac) are partially masked with diagonal lines. The source was reconstructed by ray-tracing the pixels from each image through deflection maps generated from the best-fit model, conserving the spatial resolution by allowing arbitrary pixel size and shape in the source plane. The inset in each panel shows the result of ray-tracing an image-plane point-source placed at the position of emission knot b, to illustrate the effective spatial resolution, akin to a beam pattern in a radio map. 
The excellent agreement between the four source reconstructions of the different images validates the model and helps interpret the source structure.\label{fig.sourceplane}}
\end{figure*}
\begin{figure*}
\epsscale{1.16}
\plotone{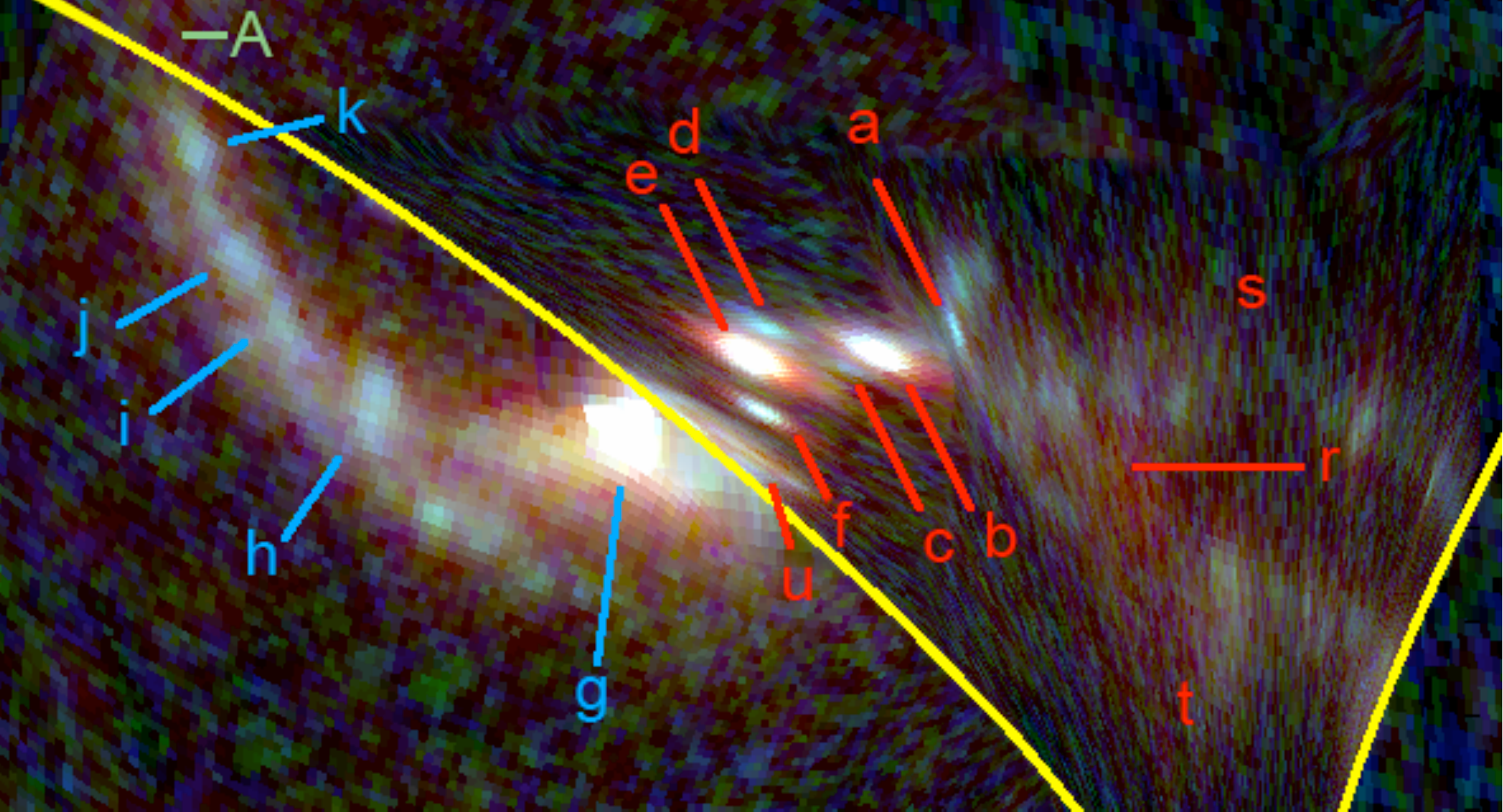}
\caption{Rendition of the source reconstruction from the four magnified images into one frame. The emission knots are labeled as in Figure~\ref{fig.1}.  \label{fig.source_labeled}}
\end{figure*}
\subsection{Source Reconstruction}\label{s.source}
The source plane reconstruction is done by ray-tracing the pixels in the observed frame through the lensing potential of the best-fit lens model (see Figures~\ref{fig.2} and \ref{fig.sourceplane}). In the process we preserve the spatial information by allowing de-lensed pixels to have arbitrary shapes and sizes. The source reconstruction is used as a tool to understand the mapping between the distorted images of the multiply-lensed source and the source itself.
The location of the source galaxy with respect to the source-plane caustics determines the geometry, multiplicity and parity of each of its lensed images. As in many cases of giant arcs, the source galaxy is positioned such that a caustic crosses it, resulting in a high magnification. In the case of RCSGA 032727-132609, as can be seen in Figure~\ref{fig.2}, the caustic that corresponds to the NE tangential critical curve separates the source into two general parts. The part that lies inside the caustic is multiply-imaged five times: its instances are images 1, 2, and part of images 3, 4 and 5. The part of the source that remains outside the caustic is multiply-imaged only three times, in images 3, 4 and 5, and ``vanishes'' in the merging pair images 1 and 2. The bright region that extends to both sides of the critical curve between images 1 and 2 is in fact mapped to an extremely small area in the source, directly inwards to the caustic. 
This source-plane to image-plane mapping may be best understood from Figure~\ref{fig.sourceplane}, which shows four separate reconstructions of the source, each one from a different lensed image. In  Figure~\ref{fig.source_labeled} (and the inset in Figure~\ref{fig.1}), we merge these four reconstructions, by combining the best-resolved segments of the source into one frame.
Notice that a faint area at the ``left'' edge of the source, which crosses back into the critical curve (designated by label A in Figure~\ref{fig.source_labeled}), is also highly magnified to form the faint tangential arc labeled as A in Figure~\ref{fig.2}.  

\begin{figure*}
\epsscale{1.16}
\plotone{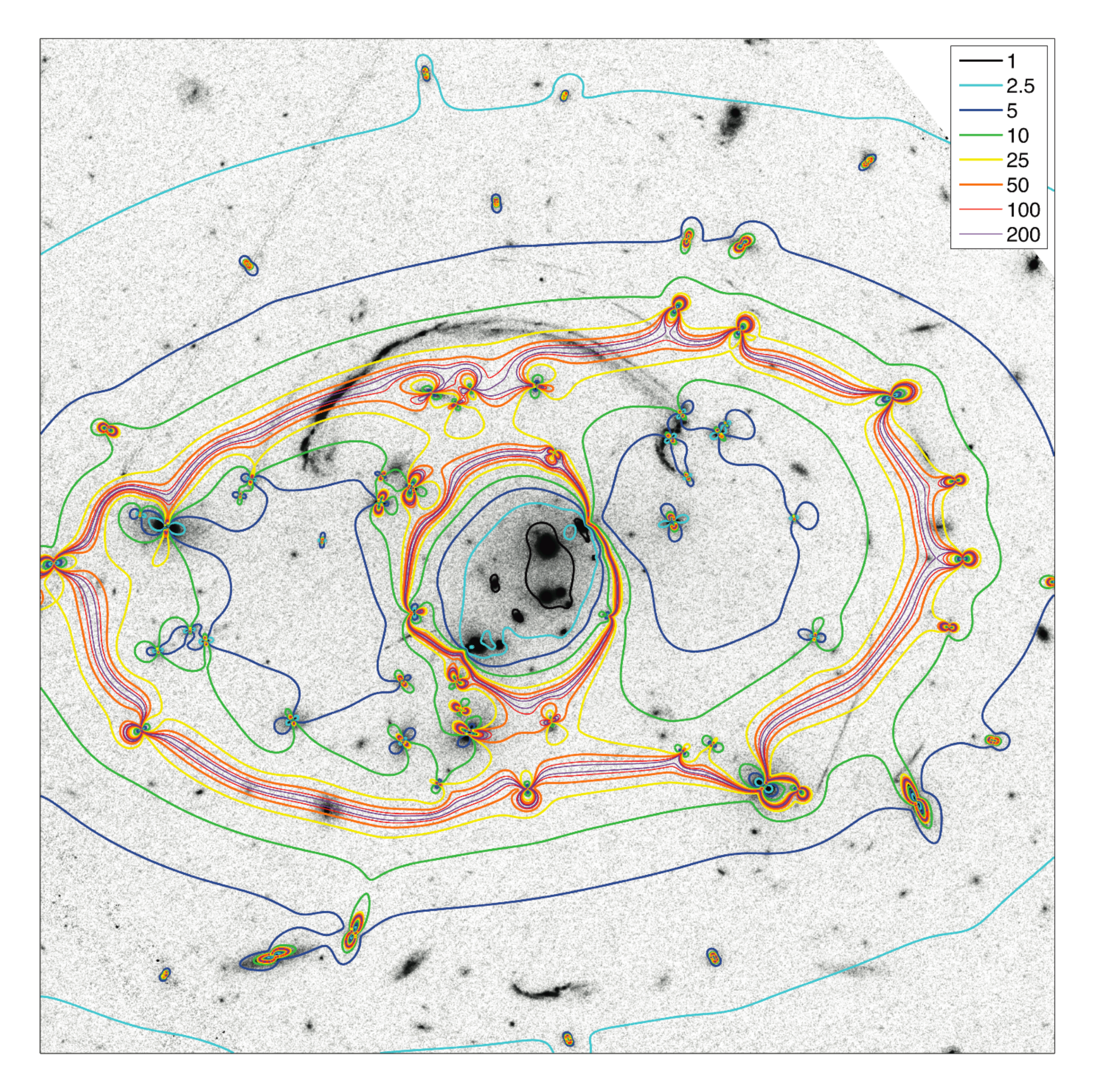}
\caption{Contours of absolute magnification for a source at $z=1.702$, overplotted on an image of the cluster field. The magnification varies across the giant arc from $\times 4$ at the western portion to $>100$ near the critical curves. The magnification range across the counter image is $2.8-2.95$. The uncertainty on the magnification is typically  $10\%$ . Magnification values between $0$ and $1$ (areas enclosed in black contours, close to the center of the cluster) mean that images that form there would be de-magnified. \label{fig.magnific2}}
\end{figure*}

\section{Magnification}\label{s.magnification}

Figure~\ref{fig.magnific2} shows image-plane magnification contours for a source at $z=1.702$, derived from the best-fit model. 
Due to the extended nature of the source, and its location in proximity to the source-plane caustics, the magnification is highly nonuniform along the giant arc, ranging from $\sim\times4$ far from the critical curves to hundreds at the critical curve. 
{We note that the effect of spatially varying magnification must be considered when interpreting measurements of the physical properties of strongly lensed galaxies, and can produce erroneous estimates of star formation rate if not accounted for (see \S~\ref{s.spec}). For example, Blain (1999) demonstrated that differential magnification of strongly lensed sub-mm galaxies can confuse measurements of the dust temperatures of those objects.}

To avoid nonlinearity and pixelization issues close to the critical curve, we measure the magnification of the arc by comparing the observed size of the image with the size of the model-reconstructed source. The magnification is the ratio of areas. 
Since some parts are highly magnified, it is instructive to measure the magnification separately for segments of the giant arc as well as the total magnification. 

We estimate the magnification uncertainty through a simulation, in which we compute many lens models, in each one drawing a set of model parameters from steps in the MCMC that are within [$\chi^2_{min}$,$\chi^2_{min}+2$]. This range encompasses the $1-\sigma$ uncertainty in the parameter space. We then measure the magnification as described above in each model, and report the magnification according to the best-fit model, and the range of magnifications found in the simulation, which is equivalent to a $1-\sigma$ confidence interval.

We find that the total magnification of the source, measured as the total area of all of its images devided by the area of the model-generated source, is \magtotal. The average magnification across the giant arc is \magA, and the counter image is magnified by \magC. The fifth image, close to the BCG, is de-magnified, i.e., it appears smaller on the sky than the source would have appeared had it not been lensed. Its magnification is \magfifth. 
Breaking the giant arc into its three separate images, we find magnifications of \magAa, \magAb, \magAc~ for images 1, 2, and 3, respectively. Note that images 1 and 2 do not represent the entire source galaxy, and therefore the total magnification of the source is not a simple linear summation of the magnifications. 
Figure~\ref{fig.longslit} shows a part of the giant arc that was targeted for longslit spectroscopy in R11, and the de-lensed location of the slit in the source plane. We find that the portion of the arc that is covered by the slit is magnified by \magslit.  

The simulations used to assess the uncertainties allow us to explore the dependence of the magnification on model parameters, and degeneracies and correlations between these parameters. Similar to Jullo et al. (2007), we find that the strongest correlation between model parameters is between $r_{core}$ and $\sigma$: higher values of $r_{core}$ result in higher values of $\sigma$. The ellipticity is anti-correlated with both $\sigma$ and $r_{core}$. A similar correlation exists between these parameters and the source magnification. Figure \ref{fig.correlations} shows the dependence of the total magnification on some of the model parameters (other parameters show no significant correlations, and are not shown). The magnification increases linearly with  $\sigma$ and $r_{core}$, and decreases with the cluster ellipticity. Interestingly, we find a correlation with the model-predicted redshift of S7, meaning that securing the redshift of this arc through spectroscopy will reduce the magnification uncertainty.

\section{The Star Formation Rate}\label{s.spec}

In R11 we used 1.3 hr of Keck/NIRSPEC spectroscopy to determine physical properties of the source, including extinction, electron temperature, oxygen abundance,and the N/O, Ne/O, and Ar/O abundance ratios. Most of these properties are independent of the lensing magnification. The exception is the star formation rate (SFR), which was measured from the flux divided by the average magnification.
This average magnification has now been better measured from our new lens model, warranting a re-examination of the star formation rate measured in R11.  Furthermore, the much higher spatial resolution of our new lensing map compared to that available to R11 enables us to better contextualize their results.

\begin{figure*}
\epsscale{1.16}
\plotone{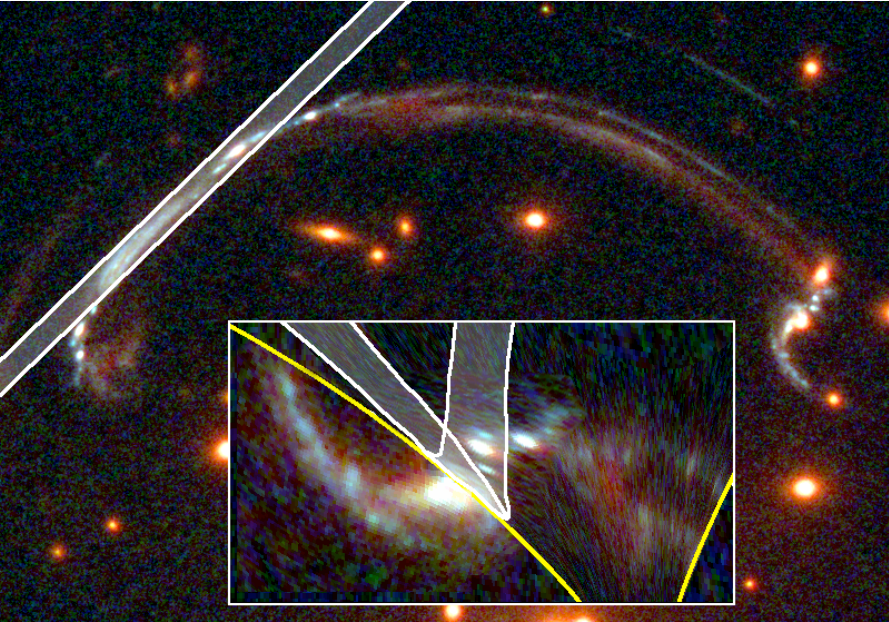}
\caption{The NIRSPEC slit from Rigby et al. (2011). The inset shows the raytracing of the slit to the source plane, using our best-fit lens model. We find that the part of the arc that is enclosed in the slit has a magnification of \magslit.\label{fig.longslit}}
\end{figure*}

\begin{figure*}
\epsscale{0.7}
\plotone{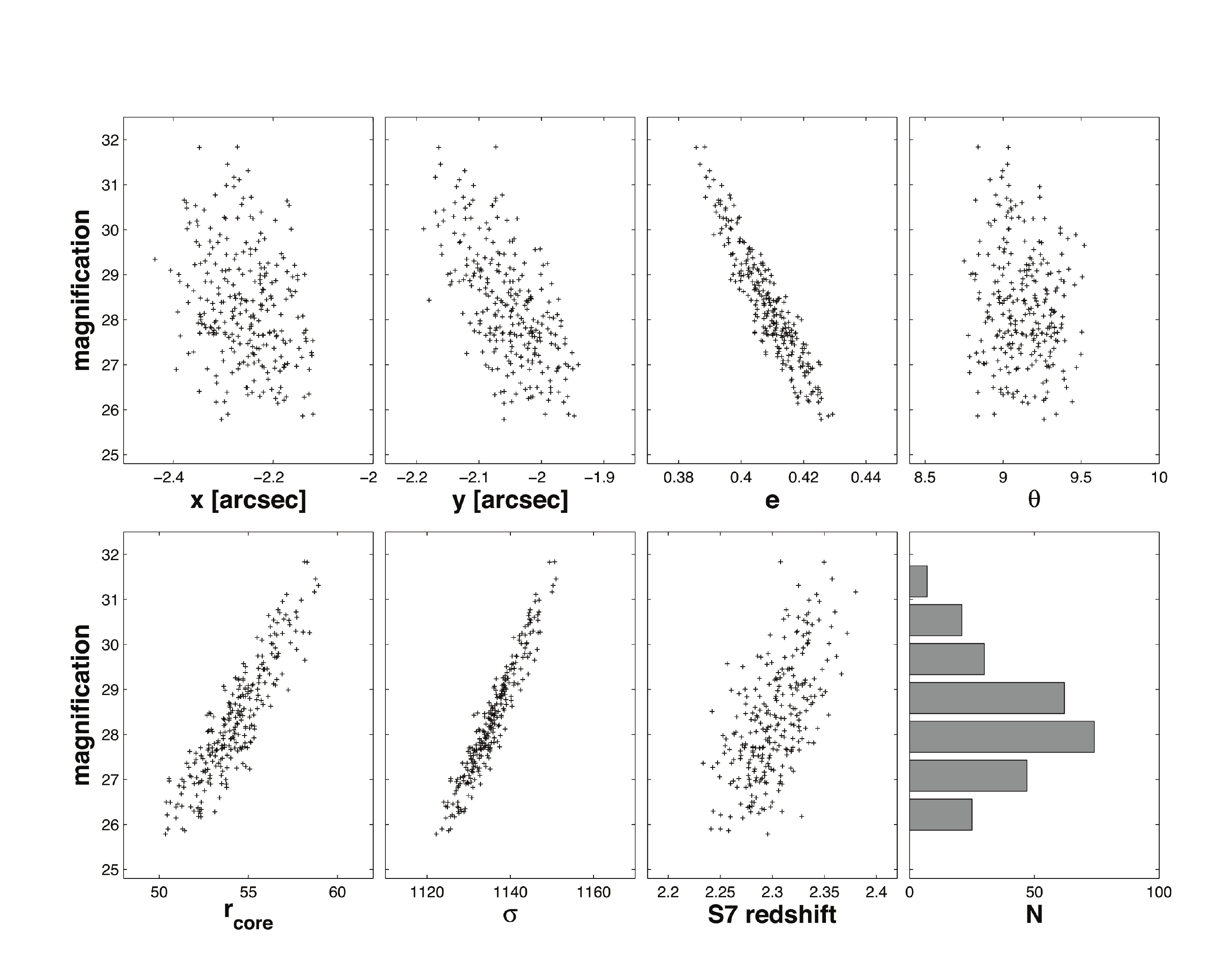}
\caption{Dependence of the total magnification on some of the model parameters, showing strong correlation with some parameters. The bottom-right panel shows the distribution of magnifications, derived from a suit of simulated models (see text).\label{fig.correlations}}
\end{figure*}

Figure~\ref{fig.longslit} shows the de-lensed R11 NIRSPEC slit on the source plane. The new lens model and source reconstruction shows that the slit targets a very small portion of the source galaxy, which is highly magnified due to its location close to the caustic. It is therefore not representative of the galaxy as a whole, but nevertheless provides a unique opportunity to probe 100 pc scale areas in the galaxy.

Wuyts et al. (2010) and R11 measured the star formation rate for \arcname~ in multiple ways:  from the broad-band optical photometry; from the 24$\mu m$ flux; and from the H$\beta$ flux measured by NIRSPEC.  They reported a discrepancy: the SFR derived from the Balmer lines is much higher than inferred from broadband photometry or 24$\mu m$ flux.   Indeed, the H$\beta$ flux measured for \arcname~ is eight times higher than was measured by Teplitz et al. (2000) for MS 1512-cB58.  

Our new lensing model solves this problem.  R11 assumed a magnification of 17.2, which was the average magnification of the entire giant arc reported by Wuyts et al. (2010).  Our new lens model shows that the portion of the galaxy captured by the NIRSPEC slit has a much higher magnification, \magslit.  The SFR inferred from H$\beta$ must thus be scaled down, to $(1170\pm60) /$ (\magslit) $=$ (\SFRinslit) $~ M_\sun {\rm yr}^{-1}$ (not corrected for extinction). This new measurement is now consistent with the SFRs inferred from the broadband SED and from the mid-IR.

\section{Summary and Conclusions}
We use new {\it HST}/WFC3 imaging data to construct a robust lens model for the lensing cluster RCS2 032727-132623. Using the new lens model, we map out the magnification of the highly-magnified arc RCSGA 032727-132609, and construct a de-lensed image of the source galaxy. Due to the location of the galaxy in the source plane with respect to the lensing caustics, parts of the source are multiply-imaged five times, while other parts are multiply-imaged only three times in a merging-pair configuration. 
In particular, we find that the brightest parts of the giant arc, which were targeted for rest-frame optical spectroscopy by R11, are lensed images of a small region of the source galaxy ($\sim 10\%$ of the spatial size of the galaxy).

\begin{figure*}
\epsscale{0.7}
\plotone{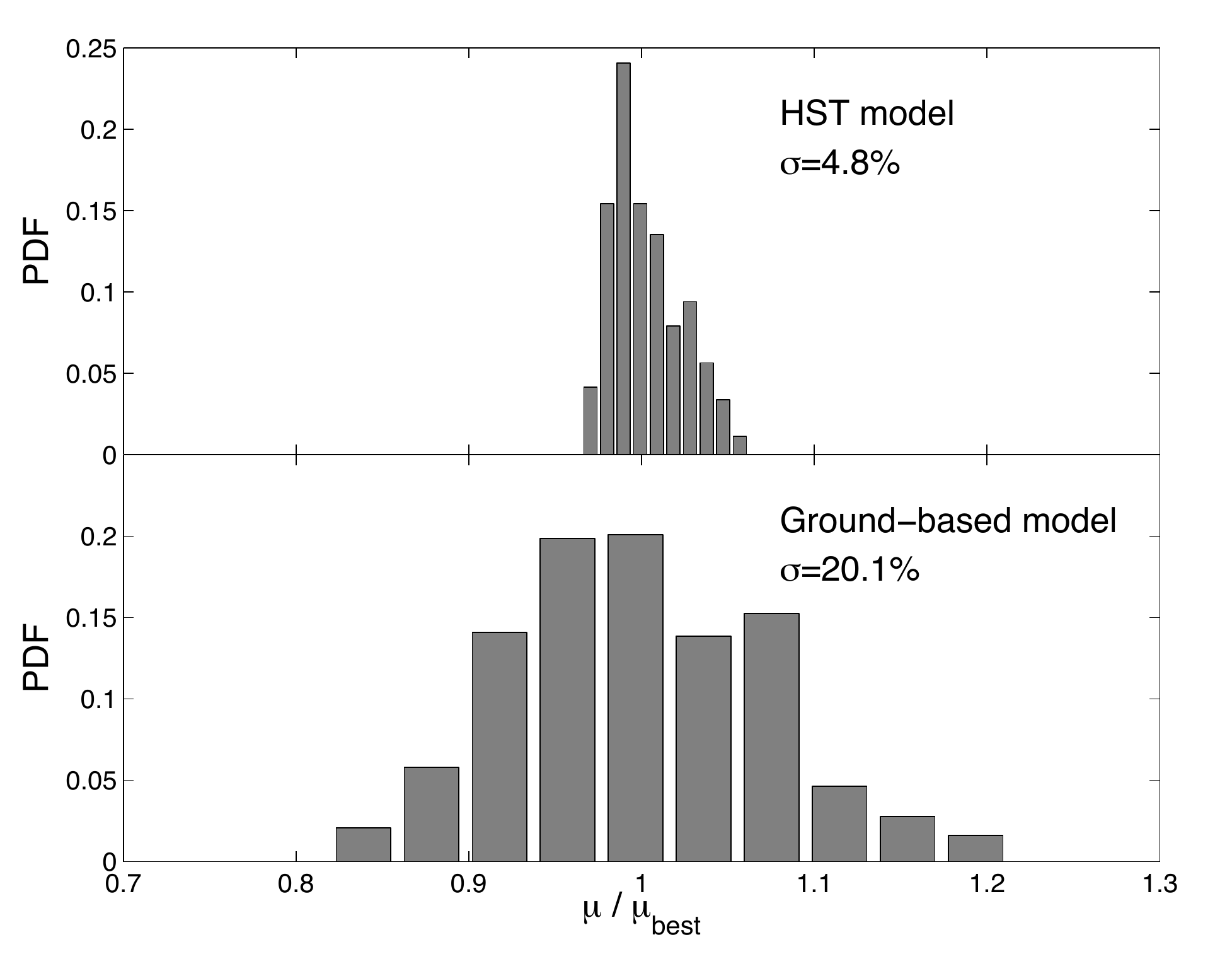}
\caption{Comparison of the magnification uncertainty between a lens model based on {\it HST} data (this work) and a lens model based on ground-based data (Wuyts et al. 2010). In each panel, we plot the {\it counter-image} magnifications, derived from a suit of simulated lens models (see \S~\ref{s.magnification}), and divided by the magnification from the best-fit model. The top panel shows the {\it HST}-based magnification histogram, and the bottom panel shows the ground-based magnification histogram. The width of each histogram represents the uncertainty in the magnification of the counter image. We find that the maginfication uncertainty  becomes four times smaller in the model based on {\it HST} data, even in the case of the counter image, whose magnification is less sensitive to the details of the model due to its distance from the critical curve. \label{fig.magnific_comp}}
\end{figure*}

The resolution of the {\it HST} data allows us to uniquely identify features in the different lensed images of the source, and identify additional lensed galaxies. The resultant lens model is superior to the previously published lens model (Wuyts et al. 2010) based on ground-based data. To quantify this claim, we compare the magnification uncertainty between models, a value that is crucial for understanding the physical conditions in the source galaxy through a measurement of the intrinsic, unlensed, luminosity (see Figure~\ref{fig.magnific_comp}). Since the ground-based lens model was not robust enough to measure the magnification of the giant arc directly, we concentrate on the easier to measure, less sensitive magnification of its counter image. In the top panel, we plot a histogram of the magnifications that were derived from a suit of simulated lens models, with parameters drawn from MCMC steps in this work. The same is shown in the bottom panel, for the model published in Wuyts et al. (2010). The magnifications in each set are divided by the magnifications from the best-fit models, to enable the comparison. We find that even in the case of the counter image, the magnification uncertainty improves by a factor of four when using high-resolution {\it HST} data. We emphasize that the model presented in Wuyts et al. (2010) is based on wrong assumptions about the mapping of the counter images to each other; in particular, the brightest emission knot marked g in Figure~\ref{fig.1} was assumed to have a counter image in images 1 (Aa) and 2 (Ab) of the giant arc, whereas we now understand that this is not the case. Thus, the magnification that was derived from it was not accurate, and the uncertainty was probably under-estimated. 

We use the new model to estimate the average magnification within the R11 NIRSPEC aperture, \magslit. Based on this magnification, we find that the average SFR within the aperture is \SFRinslit, significantly lower than the value we reported in R11, a value that was estimated under the assumption that the source is uniform and the sampled region is representative of the entire source. We remind the reader that this knowingly-wrong assumption was the only possible assumption at the time, with only a preliminary mass model in hand.
In R11, the SFR was estimated by multiplying the flux by the luminosity-weighted fraction of the galaxy that is covered by the slit, and dividing by the average magnification of the giant arc. This procedure results in an effective magnification which is several times too small, and implies an extremely high SFR for the entire source.
In practice, the region sampled by the slit, as illustrated in Figure~\ref{fig.longslit}, represents only about 10\% of the physical size of the source; it does not cover the brightest part of the source, and is probably not representative of the entire galaxy, as we suspected in R11.  
{The higher value of this updated magnification reinforces the usefulness of \arcname~ for probing down the luminosity function at high redshift, particularly with ALMA.}

The results shown in this paper allow the precise mapping of the knots in the arc to the source plane for the first time. 
This work enables many other investigations. 
In an upcoming paper we will analyze spectra on a knot-by-knot basis, to study the degree to which physical conditions vary across the arc. We also intend to use the high-resolution {\it HST} imaging presented in this paper to map on a pixel-by-pixel basis the spectral energy distribution, star formation rate, and extinction of the source, which we defer to a future paper.
Further spectroscopy should provide secure redshifts for one or more of the secondary arcs, which will reduce the uncertainty in model parameters and the magnification to the few-percent level, making \arcname~ one of the best understood lensed high-$z$ galaxies.

\begin{acknowledgements}
Support for program \#12267 was provided by NASA through a grant from the Space Telescope Science Institute, which is operated by the Association of Universities for Research in Astronomy, Inc., under NASA contract NAS 5--26555. { MDG thanks the Research Corporation for support of this work through a Cottrell Scholars award.} JRR was supported in part by a Carnegie Fellowship from the Carnegie Institute for Science. {LFB is supported by FONDECYT under project \#1085286. We thank the referee, M. Limousin, for useful comments.}
\end{acknowledgements}




\end{document}